# Orbital localization error of density functional theory in shear properties of vanadium and niobium

Y. X. Wang,[1,2] Hua Y. Geng,[1,*] Q. Wu,[1] and Xiang R. Chen[3]

[1]National Key Laboratory of Shock Wave and Detonation Physics, Institute of Fluid Physics, CAEP, Mianyang 621900, Sichuan, People's Republic of China

[2]College of Science, Xi'an University of Science and Technology, Xi'an 710054, People's Republic of China

[3]Institute of Atomic and Molecular Physics, College of Physical Science and Technology, Sichuan University, Chengdu 610064, China

## Abstract

It is believed that the density functional theory (DFT) describes most elements with *s*, *p* and *d* orbitals very well, except some materials that having strongly localized and correlated valence electrons. In this work, we find that the widely employed exchange-correlation (xc) functionals, including LDA, GGA and meta-GGA, underestimate the shear modulus and phase stability of V and Nb greatly. The advanced hybrid functional that is usually better for correlated system, on the other hand, completely fails in these two simple metals. This striking failure is revealed due to the orbital localization error in GGA, which is further deteriorated by hybrid functionals. This observation is corroborated by a similar failure of DFT+*U* and van der Waals functionals when applied to V and Nb. To remedy this problem, an semi-empirical approach of DFT+*J* is proposed which can delocalize electrons by facilitating the on-site exchange. Furthermore, it is observed that including density derivatives slightly improves the performance of the semi-local functionals, with meta-GGA outperforms GGA, and the latter is better than LDA. This discovery indicates the possibility and necessity to include higher-order density derivatives beyond the Laplacian level for the purpose to remove the orbital localization error (mainly from *d* orbitals) and delocalization error (mainly from *s* and *p* orbitals) completely in V and Nb, so that to achieve a better description of their electronic structures. The same strategy can be applied to other *d* electron system and *f* electron system.

---

*Correspondence and requests for materials should be addressed to H.-Y.G. (email: s102genghy@caep.ac.cn).

# I. INTRODUCTION

Quantum mechanics computation is indispensable for modern condensed matter physics and materials design, in which DFT, owing to its high efficiency and acceptable accuracy, has become one of the most widely employed methods for many-body electronic structure calculations.[1-4] With appropriate approximation in the exchange-correlation (xc) functional, DFT has successfully been applied to a variety of important problems across wide disciplinary fields.[5-11]

Nonetheless, DFT suffers for lacking a strategy to improve its accuracy systematically. In practice, variants of functionals have been proposed, which usually were benchmarked against experimental data and/or other more accurate quantum mechanics calculations, such as quantum Monte Carlo method, configuration interaction, and coupled-cluster theory. One of the well-known problems in DFT is the underestimation of the band gap in insulators and semiconductors,[12-14] as well as the underestimation of the on-site electronic correlations in narrow band systems (especially those involve localized $d$ or $f$ valence electrons).[15,16] This deficiency closely relates to the intrinsic self-interaction error[17] and the consequent *delocalization error* in the Kohn-Sham framework.[18,19]

Note that the terms of "localization error" and "delocalization error" as standard definition are to mean super-linear (concave downward) or sub-linear (concave upward) variation of the total energy in an isolated open system as a function of electron number between two adjacent integers. The exact variation should be linear. Semi-local functionals applied to finite systems usually show delocalization error, whereas Hartree-Fock (HF) theory applied to such systems shows the localization error. In a bulk system, any approximate functional that predicts it to be a metal will have the correct linear behavior. Nonetheless, approximate functional often leads to density-driven error even in metals, where some orbitals could be too delocalized or too localized. The resultant error can be termed as *orbital localization error* and *orbital delocalization error*. They are a generalization of the standard definition to an extended system.

To remedy the delocalization error, hybrid functional[20] was proposed, which includes a fraction of HF type exact exchange into the LDA[2] and GGA[13] functionals. Application of this method demonstrated that hybrid functionals are essential to improve the DFT performance: the band gap, binding energy, and structural geometry all are systematically improved with respect to the local and semi-local functionals.[21,22] It now becomes the state-of-the-art approach for modern DFT applications and xc functional development.[20,23-25]

Even though transition-metal clusters and compounds usually have strong electronic correlations, an appropriate treatment of which requires the hybrid functional or DFT+$U$ method, it is long believed that bulk transition metals can be described by pure DFT very well,[26-30] even in the local and semi-local xc functional level, because of the open-shell electronic configuration dominated by the well delocalized $s$ and $d$ electrons. Previous investigations indeed illustrated that the PBE parameterization of GGA functional provides a good description of the equation of state (EOS) for the group VB metal V[31] and Nb.[32] The predicted body-centered cubic (BCC) to rhombohedral (RH) transition in V[33-35] was also confirmed by DAC experiment.[36] Other properties of V and Nb, such as electron-phonon coupling,[37,38] superconducting properties,[39,40] and melting properties[41] were also described well by PBE.

Unfortunately, PBE greatly underestimates the shear modulus $C_{44}$ by a magnitude of 40% and 30% for V and Nb at 0 GPa, respectively, even though other elastic moduli $C_{11}$ and $C_{12}$ are well reproduced.[42] It is exceptional if compared to Al and Li whose shear modulus were perfectly reproduced by PBE, regardless of their much smaller magnitude of the shear modulus. On the other hand, PBE correctly captures the pressure-induced $C_{44}$ softening that closely relates to the strong electron-phonon coupling due to the Kohn anomaly of the Fermi surface nesting in V and Nb.[42-44] The major contribution of this electronic effect has been automatically taken into account when one traces the energy variation along the shear deformation path when computing the elastic modulus. Unless there is a strong violation of the *classical* Born-Oppenheimer approximation, which is *unlikely* by considering the large mass of these two elements, there is no reason to impute the observed $C_{44}$ underestimation to the nesting structure in the Fermi surface (FS) rather than the approximate xc functional[42-44] (note that Li also has FS nesting, but its shear modulus is well described by PBE). From another perspective, if the deviation in $C_{44}$ is purely because of the FS feature and the consequent electron-phonon coupling, then the deviation should independent of the xc functionals. Unfortunately, as will be seen below, it is not true. In fact, the quality of the xc functional is the most important factor that affects the result.

Another indication of the limitation of the PBE functional in vanadium is that, though the predicted BCC→RH transition is qualitatively in agreement with the experimental observation, the discrepancy on the transition pressure becomes unexpectedly larger if a careful calculation with good numerical convergence is carried out.[35,45] Considering the fact that the semi-local functional such as PBE might be incompetent in dealing with the strong electronic correlations in narrow band systems (such as actinides, lanthanides, and some transition metals and their compounds),[15,16] most of its error can be corrected by hybrid functionals.[21,22] It is reasonable to expect a similar improvement in V and Nb. Therefore, in order to correct the errors in PBE as mentioned above and to improve the performance of DFT, a good choice should be to go beyond PBE level and employ the hybrid functionals.

In this work, the performance of widely used xc functionals on the phase stability and shear properties of V and Nb is thoroughly evaluated. An unanticipated deficiency of DFT (especially the hybrid functional) is discovered. This failure is traced back to a very rare error of DFT in bulk metals, *i.e.*, the *localization error*, which has long been ignored.[18] Our results suggest that local and semi-local functionals such as LDA and GGA can have both localization and delocalization error simultaneously. In next section, the calculation method is briefly presented. The main results and discussion are given in Sec. III and IV, with the conclusions are summarized in Sec. V.

## II. METHODOLOGY

All simulations reported in this work were performed using the Vienna *Ab initio* Simulation Package (VASP),[46] which is based on DFT and the projector augmented-wave (PAW) method.[47] The PAW pseudopotential contains 13 valence electrons (including $4s^2$, $4p^6$, $4d^4$, and $5s^1$ states). The kinetic energy cut-off of the plane waves is taken as 900 eV. In order to evaluate the performance of various xc functionals on the elastic modulus $C_{44}$ of V and Nb thoroughly, besides the widely employed LDA and GGA functionals, other four different types of advanced correction to the xc functional are also adopted: (*i*) Hybrid functionals with an appropriate mixing between the density functional and Hartree-Fock theory, such as HSE06,[48-50] PBE0,[51,52] and B3LYP;[53,54] (*ii*) DFT+*U* (using PBE xc functional, *i.e.*, PBE+*U*) with inclusion of the on-site Coulomb repulsion and exchange parameters *U* and *J*, following the recipe of Liechtenstein *et al.*;[55] (*iii*) Meta-GGA with inclusion of the second derivative of the electron density (the Laplacian), namely, the third rung of "Jacob's ladder" for functional approximations, above the LDA and GGA rungs. This includes TPSS, revised-TPSS (RTPSS)[56] and SCAN;[57] (*iv*) Van der Waals functionals,[58] an approximation of interacting response functions in terms of electronic density to account for nonlocal dispersion correlation energy, which includes optB86b-vdW, optPBE-vdW, revPBE-vdW, and vdW-DF2. For the *k*-point sampling in the Brillouin zone, a $40\times40\times40$ uniform mesh are used for the calculation of LDA, PBE, TPSS, RTPSS, PBE+*U*, and all van der Waals functionals, whereas a $14\times14\times14$ uniform mesh is used for the hybrid functionals HSE06, PBE0, and B3LYP. This parameter setting is carefully checked by increasing the cutoff energy and *k*-points to higher values to ensure an absolutely converged total energy.

### III. RESULTS

### A. Equation of state

DFT is exact in principle. If exact xc functional were used, DFT should have correctly described the structure and mechanics of condensed matter.[59] From this point, one expects that hybrid functionals that include portion of the exact exchange contribution should give a better shear modulus $C_{44}$ for V and Nb. We will present a thorough and comprehensive discussion on this topic below. Before moving to this mechanical property, it is necessary to assess the performance of different DFT functionals on the equations of state (EOS) of V and Nb under high pressures.

This calculation covers almost all commonly used xc functionals that are available in most DFT codes. The main EOS results, together with the reduced shock wave experimental data for vanadium,[60] are shown in Fig. 1. It is evident that most functionals underestimate the equilibrium atomic volumes at zero pressure, whereas vdW-DF2 slightly overestimates it. Only revPBE-vdW predicts a density that matches the experimental data perfectly at low pressures. Except the LDA, which underestimates the atomic volume by about 1.3 Å$^3$/atom, and the meta-GGA and hybrid functionals that have an underestimation of about 0.7 Å$^3$/atom, all other functionals predict a volume deviation less than 5% at 0 GPa. This deviation reduces with increasing pressure, and is about 4% at 100 GPa, suggesting the overall improvement of the DFT performance at higher pressures. Within the explored pressure range, our calculated EOS does not show any kinks, and is in good agreement with previous theoretical study.[43] This validates our computational method and the parameter setting.

### B. Shear modulus from DFT

Different from the EOS that is trivial, an intriguing anomaly takes place in shear modulus. Our calculated $C_{44}$ for V and Nb with various DFT functionals are summarized in Table I. Other available theoretical results[42,44,61] and experimental data[62-69] are also included for comparison. It clearly shows that the $C_{44}$ calculated by GGA (in PBE) is significantly underestimated by about -40% (-30%) for V (Nb) with respect to the experimental data. Most seriously, the $C_{44}$ calculated by all hybrid functionals, including PBE0, HSE06, and B3LYP, are even negative at zero pressure. This demonstrates that all hybrid functionals incorrectly predict a mechanical instability of the BCC structure for both V and Nb. To the best of our knowledge, this unphysical performance of the hybrid functional has not been aware of before.

Similarly, the $C_{44}$ of LDA and meta-GGA (both TPSS and RTPSS) are greatly underestimated by about -71% and -45% for V and -59% and -23% for Nb, respectively, compared to the experimental data. Meanwhile, the recently developed meta-GGA functional SCAN[57] has a deviations in $C_{44}$ by about -44% and -26% for V and Nb, respectively. As shown in Table I, the result of meta-GGA is a little bit better than GGA and LDA for Nb. The van der Waals functionals (vdW-DF), which is usually important for soft matter and closed shell electronic system, also underestimate the shear modulus by about -50% for both V and Nb. In this sense, all local and semi-local functionals outperform the non-local and hybrid functionals on this important issue.

In Table II, we list the calculated elastic moduli $C_{11}$ and $C_{12}$, bulk modulus $B_0$ and its pressure derivative $B_0'$ for V at zero pressure by using different DFT functionals, together with the available experimental[62-67,69] and theoretical data.[70] It can be seen that our calculated $C_{11}$, $C_{12}$, $B_0$ and $B_0'$ by using different functionals are consistent with each other, and in good agreement with the experimental results. Based on this we can conclude that the monoclinic shear deformation and its induced mechanical instability are much more sensitive than other properties to the accuracy of the xc functional. It thus provides a benchmark that is better than the lattice constant, bulk modulus, and EOS that can be employed to calibrate the xc functionals.

On the other hand, it had well been established that at ambient and low pressures, both V and Nb are in a simple BCC structure. We take the BCC structure as the starting point, then compute its energy variation along different deformation path. The calculated energy difference of vanadium with respect to the BCC phase as a function of the rhombohedral distortion magnitude $\delta$[45] by using various xc functionals at 0 GPa are plotted in Fig. 2. It is evident that the BCC structure is the ground state for LDA, GGA, and TPSS functionals. This agrees with the experimental observations.[31,71] Nonetheless, hybrid functional PBE0, HSE06, and B3LYP all predict that the RH$_1$ phase is the most stable structure, with BCC becomes thermodynamically and dynamically unstable. This unphysical result is in line with the predicted negative $C_{44}$ as mentioned above, showing that hybrid functional wrongly describes the phase stability of V. The same results are also obtained for Nb. In hybrid functional, after inclusion of the exact exchange contribution, the stability of high-symmetric BCC phase is completely destroyed. Its energy landscape becomes a double well, with BCC at the saddle point, and $RH_1$ and $RH_2$ becomes the global and local minimum in energy, respectively. The energy difference with respect to the BCC structure at $\delta$ = -0.06 is -0.048, -0.061, -0.071 eV/atom (and -0.013, -0.021, -0.039 eV/atom at $\delta$ = 0.06) for HSE06, PBE0, and B3LYP, respectively. This observation clearly manifests why hybrid functionals predict a negative shear modulus. The large underestimation of $C_{44}$ by local and semi-local functionals is also owing to the flat variation of the energy surface around BCC.

### C. DFT+$U$ and localization error

An interesting question is what causes this failure of the hybrid functional? Considering the successful application of hybrid functionals on a wide range of problems and materials,[21,22] it becomes necessary and imperative to understand the origin of this apparent failure, which might be helpful for devising a better functional. It is necessary to point out that the variation of the energy as shown in Fig. 2 is very similar to the pressure-induced variation in vanadium[30] and the change in niobium by removing some valence electrons.[27] This suggests hybrid functional might have incorrectly enhanced the nesting feature in the Fermi surface, and effectively reduced the number of valence electrons. On the other hand, according to the construction of the hybrid functional, it is easy to figure out that its unphysical behavior should stem from the exact exchange part of the HF contribution. It is well known that HF theory suffers from the localization error in a finite system, and overestimates the band gap.[66] This HF contribution, because of having included the exact exchange, is helpful in correcting the self-interaction error that is intrinsic to DFT. This is the main reason to combine it with the LDA (or GGA), in a hope that its localization error and the delocalization error of the latter can *cancel out* and then improve the performance of the resultant hybrid functional. In order to verify this hypothesis, we focus on the orbital localization error of the HF part. Our approach is to compare the results with another method, DFT+$U$, which uses the Hubbard model to enhance the orbital localization. By adjusting the effective coulombic repulsion parameter $U$ that acting on the $d$ orbitals, one can tune the $d$ electrons in the system evolving from a delocalized valence state into a highly localized state. For the DFT part, we use the PBE functional. Therefore it corresponds to PBE+$U$.

The evolution of the phase stability of vanadium with the on-site coulomb repulsive strength $U$ is similar to that of hybrid functional, as shown in Fig. 3. The enhanced localization of $d$ electrons first flattens the energy surface, leading to a reduced shear modulus. After $U \geq 3$ eV, a double well develops, which destroys the stability of the BCC phase. The RH phases become preferred, and results in a negative shear modulus in BCC structure. These observations are exactly the same as that observed in hybrid functionals, except here $RH_2$ is slightly favored over the $RH_1$.

This observation is further corroborated by the explicit comparison of the localization degree of electrons. We compute the differential electron localization function (ELF) and the differential charge density of PBE+$U$ ($U$=4 eV) and the hybrid functional HSE06 with respect to PBE, respectively. The obtained results are shown in Fig. 4. It is evident that both of them attract electrons (*i.e.*, negatively charged) and localize them around the nuclei when compared to the PBE results. This is a strong evidence for the similarity of the electron density distribution between the hybrid functional and the PBE+$U$ method, and confirms that the failure of hybrid functionals in V and Nb is owing to the orbital localization error that is worsened by the HF contribution.

Back to the failure of the LDA and GGA functionals. It is well-known that these functionals have severe orbital delocalization error, and underestimate the band gap in insulators and semiconductors. On the other hand, compared to the hybrid functional results, the large underestimation of the shear modulus in V and Nb suggests that even without the HF contribution, LDA and GGA still have some orbital localization error. This observation of *simultaneous* localization and delocalization error in DFT is important, and requires a careful assessment. According to the construction nature of LDA, one cannot expect it to have a big localization error. By analysing the variation of potential energy surface (as well as the binding strength of the BCC structure and the value of $C_{44}$) with the available number of valence electrons by using the method of charge transfer and jellium model for V[35] and Nb,[32] we find that it seems the delocalization error in LDA should be larger than the localization one. On the opposite, the localization degree in PBE might be too larger, which also weakens the stability of BCC, just from another side. This argument is supported by the result of vdW functionals, which enhance the electron localization by comparing to PBE. Along this route, in order to remedy this difficulty, one would like to introduce the contribution of higher orders of the density derivatives. Meta-GGA is one of that including the Laplacian of the density.

As shown in Fig. 2, the TPSS functional of meta-GGA captures the phase stability of V and Nb at zero pressure correctly. Specifically, TPSS and RTPSS underestimate the $C_{44}$ by about -47% and -44% for V, respectively, which is similar to PBE, and is much better than other functionals, as listed in Table I. For Nb, the correction of meta-GGA is more obvious, and the calculated $C_{44}$ of TPSS and RTPSS are 25.26 and 24.24 GPa, respectively, very close to the experimental data of 28.7 GPa.[69] It is obvious that meta-GGA improves the DFT performance by comparing to pure GGA. Nonetheless, it is still short of accounting for the whole discrepancy between the DFT and the experimental data in V. It seems including higher orders of the density derivatives might be helpful for solving this problem. Unfortunately, such a functional is not available yet.

### D. Correction of the localization error---DFT+$J$

In order to correct the orbital localization error in V and Nb as mentioned above, a tentative semi-empirical Hubbard model is proposed in this work. This approach is similar to the DFT+$U$ method and can be called the DFT+$J$ method, where the parameter $J$ represents an equivalent kinetic term allowing for tunneling (hopping) of electrons. By adjusting the effective exchange parameter $J$ that acting on the $d$ orbitals, one can increase the on-site exchange probability of the $d$ electrons in the system. In order to evaluate the performance of PBE+$J$ method on this challenging problem, we first calculate the phase stability of V for varying values of $J$ at 0 GPa, with the results are presented in Fig. 3. As expected, we find that the BCC phase becomes more favored when $J$ is increased. This behavior is contrary to the PBE+$U$ method. We also find that the contribution of this model-enhanced exchange is non-linear. It reaches the maximum when $J$=2 eV ($J$=1 eV) where the BCC structure attains its greatest stability for V (Nb). Beyond this value the energy decreases, and the stability of BCC is weakened (see supplementary material Figs. S1-S2). The calculated $C_{44}$ with the PBE+$J$ method is also listed in Table I. As shown, this method corrects the shear modulus $C_{44}$ from 25.5 to 37.34 GPa for V with $J$=2 eV (and from 19.77 to 22.07 GPa for Nb with $J$=1 eV). We should note that the choice of the value of $J$ is empirical. But by using this approach, the $C_{44}$ deviation in vanadium from the experimental data is corrected to less than 12% (or 8% if RTPSS+$J$ is used). The performance on other elastic moduli and EOS is also good.

The good performance of PBE+$J$ method can be understood by comparing its electron localization function (ELF) and charge density distribution to that of pure PBE. The calculated differential ELF and differential charge density between PBE+$J$ (with $J$=2 eV) and PBE are shown in Fig. 4 for V. The feature is quite different from that of PBE+$U$ or HSE06. Now the electrons are expelled away from the nuclei (*i.e.*, positively charged), and become less localized. This confirms that DFT+$J$ method indeed can correct the localization error in semi-local functionals such as GGA and meta-GGA. It also demonstrates that the localization error is mainly from the localized $d$ orbitals. To verify this, we also apply the PBE+$J$ (with $J$=2 eV) method to the $s$ (or $p$) orbital. The results show that the $C_{44}$ is degraded by about 4.97 GPa or 19.5% (17.97 GPa or 29.5%) by comparing to the pure PBE, revealing that there has some additional delocalization error in the $s$ and $p$ orbitals. On the other hand, applying hybrid functional or DFT+$U$ to the $s$ or $p$ orbitals enhances the BCC stability in V and Nb.

Above results and discussion unequivocally demonstrate that: (*i*) the hybrid functionals and DFT+$U$ are completely failed for the phase stability and shear properties of V and Nb; (*ii*) LDA, GGA, and vdW functionals all underestimate the shear modulus $C_{44}$ and the BCC stability of V and Nb greatly, in which PBE is the best one; (*iii*) including higher orders of the density derivatives (meta-GGA) improves the performance of the xc functional, but is not enough to account for the whole discrepancy with respect to the experimental data; (*iv*) the semi-empirical DFT+*J* approach can correct the *orbital localization error* of semi-local functionals, and well reproduces the experimental shear modulus.

## IV. DISCUSSION

One further important question is how this correction affects the high-pressure properties? For hydrostatically compressed V and Nb up to 300 GPa, we calculate the elastic moduli $C_{11}$, $C_{12}$, and $C_{44}$ at zero Kelvin as a function of pressure by using different xc functionals and the PBE+*J* method. The results of vanadium are shown in Fig. 5. It is evident that the calculated $C_{11}$ and $C_{12}$ by different methods are consistent well with each other, indicating that they are insensitive to the localization and delocalization error. However, the $C_{44}$ as a function of pressure calculated by different methods is quite different. The general feature is that all corrections upshift the calculated shear modulus almost in parallel, with PBE+*J* has the largest correction. The most interesting thing is both meta-GGA and PBE+*J* increases the BCC→RH$_1$ transition pressure. Especially, the $C_{44}$ of PBE+*J* becomes positive within the studied pressure range, indicating that the correction of localization error weakens (or even eliminates) the RH phases. It is worth mentioning that the $C_{44}$ of V gradually increases along the Jacob's ladder of PBE→meta-GGA→PBE+*J*→ meta-GGA+*J* at a given pressure, suggesting that the exact DFT functional should also have similar behavior on the phase stability of vanadium.

This conclusion is intriguing. High-pressure experiment with DAC showed that the BCC→RH$_1$ transition takes place at about 63-69 GPa.[31] But the observed structure change is very subtle, and is difficult to identify in x-ray diffraction experiment. A recent independent DAC experiment reported a lower transition pressure of 30 GPa,[72] revealing the large uncertainty in the experiments. On the other hand, though a PBE calculation with poor convergence can give a transition pressure as low as 32 GPa,[35,73] the well converged PBE results is higher than 84 GPa.[45] Here, with a method that is better than PBE and has corrected the localization error, we find the transition pressure is increased further to about 118 GPa in TPSS (and higher in RTPSS). Even more, PBE+$J$ (and meta-GGA+$J$) completely eliminates the stability range of the RH phases in vanadium, challenging the previously reported structural transition. It cannot be excluded that what was observed in DAC experiments could be just the RH distortions induced by deviatoric stress, rather than a true new phase (this is supported by the small change in the $\alpha$ angle as observed in the experiment). In this respect, the BCC→RH transition in high-pressure vanadium is still an open question both in experiment and in theory. Its ultimate solution requires an accurate treatment of the orbital localization and delocalization errors in DFT simultaneously.

## V. CONCLUSION

In summary, a thorough and comprehensive assessment of the xc functionals in transition metal V and Nb was carried out. It is found that the widely used functionals, including LDA, GGA, meta-GGA, and all variants of vdW functionals, greatly underestimate the shear modulus $C_{44}$ and the mechanical stability of BCC structure of V and Nb. The advanced hybrid functionals, including HSE06, PBE0, and B3LYP, completely fail on these two metals. A systematic analysis traced this unexpected failure back to the localization error in semi-local functionals, which is further worsened by hybrid functionals. This conclusion was corroborated and supported by a similar failure of DFT+$U$ method. In order to correct this localization error, a DFT+$J$ method was tentatively proposed, which effectively facilitates on-site electron exchange, and reproduces a good shear modulus and phase stability for V and Nb. It should be noted that the DFT+$J$ is a semi-empirical approach, and the value of $J$ requires experimental data or other accurate quantum mechanics results as a reference. This method is recommended only when the calculated elastic modulus of the DFT deviates from the experimental data by more than 10%. The same strategy can be applied to other systems containing $d$ electrons and $f$ electrons.

Our investigation also revealed that including higher orders of the density derivatives improves the performance of the xc functional, with the meta-GGA outperforms PBE and LDA. This showcases a possible strategy to eliminate the *simultaneous localization error in d orbitals and delocalization error in s and p orbitals* by including the contribution of higher orders of density derivatives beyond the Laplacian level, a route towards an exact density functional.

Finally, we should note that the error in DFT can also be alternatively separated into functional-driven and density-driven errors, in which the latter can be reduced or eliminated by using an accurate density as input.[74] Here we did not perform this error-analysis because of the difficulty to obtain an accurate density for metallic solid. This, however, will not change the main conclusions obtained in this work.

**Acknowledgments**

This work was supported by the NSAF under Grant No. U1730248 and U1830101, the National Natural Science Foundation of China under Grant Nos. 11672274, 11602251, and 11904282, the Fund of National Key Laboratory of Shock Wave and Denotation Physics of China under Grant Nos. 6142A03010101 and JCKYS2018212012, the CAEP Research Project under Grant No. CX2019002, and the Science Challenge Project Tz2016001. H.Y.G appreciates Prof. J. P. Perdew for helpful discussion.

**Competing Interests**

The authors declare no conflict of interest.

**Data Availability**

All data generated or analyzed during this study are included in this published article (and its supporting information file).

**Author Contributions**

H.Y.G. conceived and designed the research, Y.X.W. and H.Y.G. performed the research, all authors analyzed the data and wrote the paper.

**Supplementary Material**

Computational details, employed xc functionals introduction, additional table and figures.

**Table I.** Calculated shear modulus $C_{44}$ of V and Nb at the ambient pressure by using different xc functionals, including the PBE+$J$ method ($J$=2 eV for V, and $J$=1 eV for Nb), together with available experimental and other theoretical data.

| $C_{44}$ (GPa) | GGA-PBE | LDA | TPSS | RTPSS | SCAN | optB86b-vdW | PBE0 | HSE06 | B3LYP | TPSS+$J$ | RTPSS+$J$ | PBE+$J$ | Expt. |
|---|---|---|---|---|---|---|---|---|---|---|---|---|---|
| V | 25.5 17.1[44] 24[61] | 12.2 | 22.4 | 23.7 | 23.6 | 18.3 | -89.9 | -29.8 | -10.2 | 36.8 Δ=13% | 39.1 Δ=8% | 37.3 Δ=12% | 42.5-44 [62-68] |
| Nb | 19.7 10.3[44] 17.2[42] | 11.8 | 25.3 | 24.2 | 21.3 | 16.3 | -66.4 | -7.79 | -3.1 | 26.8 Δ=6% | 25.6 Δ=11% | 21.2 Δ=26% | 28.7[69] |

**Table II.** Calculated elastic moduli $C_{11}$ and $C_{12}$, bulk modulus $B_0$ (GPa) and its pressure derivative $B_0'$ of V by using different DFT functionals and PBE+$J$ method ($J$=2 eV) at zero pressure, together with the available theoretical and experimental data.

| 0 GPa | $B_0$ (GPa) This work | $B_0$ (GPa) Theor.[70] | $B_0'$ | $C_{11}$ (GPa) | $C_{12}$ (GPa) |
|---|---|---|---|---|---|
| PBE | 181.25 | 183.1 | 3.85 | 266.82 | 138.47 |
| HSE06 | 196.11 | 214.9 | 3.19 | 258.95 | 164.69 |
| TPSS | 194.55 | 198.9 | 3.97 | 276.51 | 153.54 |
| RTPSS | 198.10 | 212.0 | 3.92 | 280.49 | 156.91 |
| PBE+$J$ | 187.85 | | 4.01 | 279.69 | 141.82 |
| optB86b-vdW | 190.97 | | 3.99 | 273.85 | 149.53 |
| optPBE-vdW | 179.51 | | 3.94 | 261.74 | 138.39 |
| revPBE-vdW | 169.52 | | 3.74 | 251.18 | 128.69 |
| vdW-DF2 | 167.64 | | 3.72 | 250.92 | 125.99 |
| Expt.[62-67] | — | | — | 227.9-232.2 | 118.8-121.0 |
| Expt.[69] | 195.30 | | 3.52 | — | — |

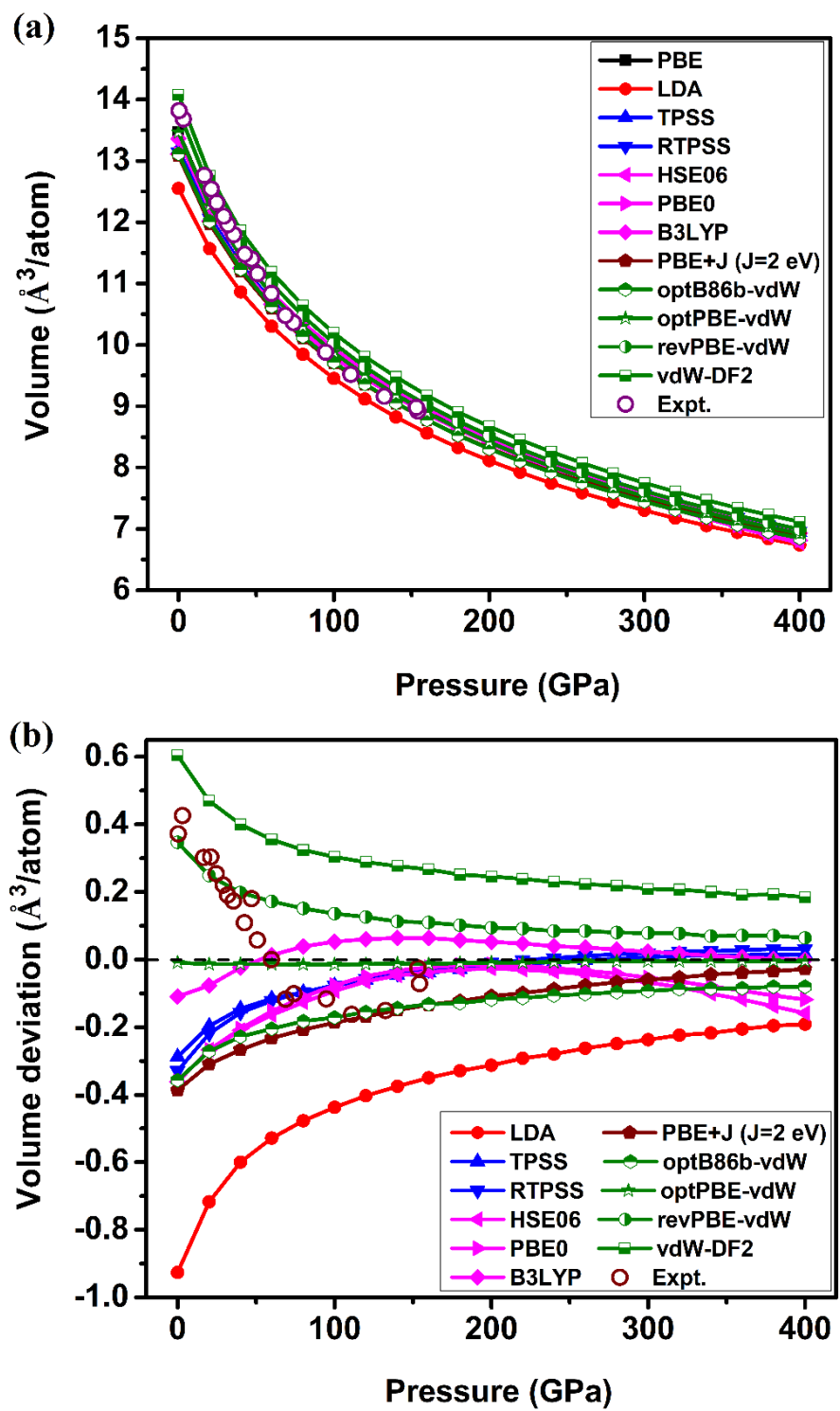

**Figure 1.** (Color online) (a) Calculated isothermal EOS (P-V curve) of V at zero Kelvin. The open circles are the experimental data taken from Ref. 60. (b) The deviation of the atomic volume with respect to the PBE results as a function of pressure.

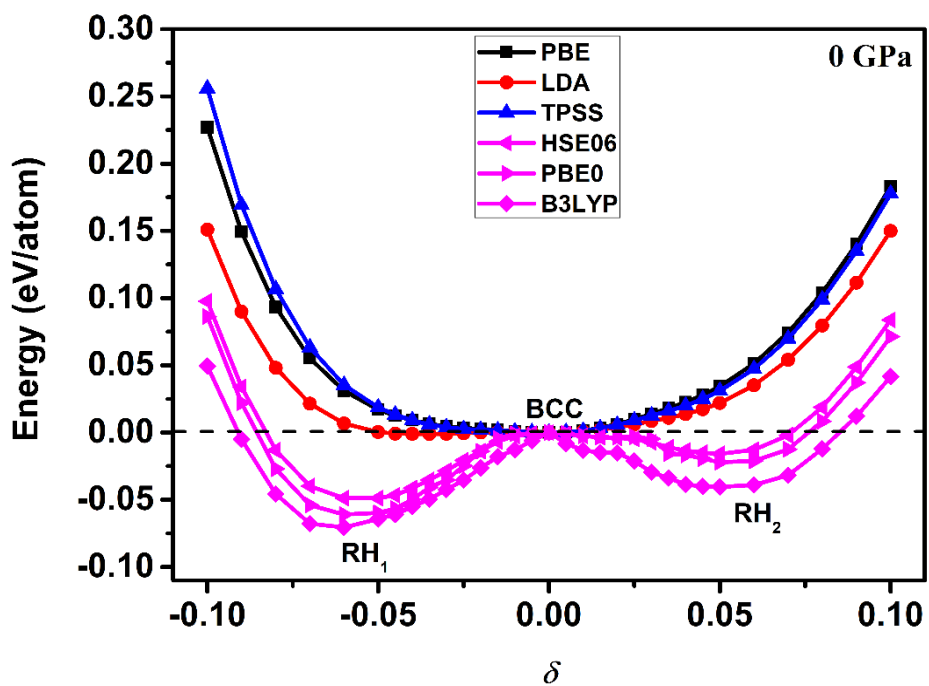

**Figure 2.** (Color online) Variation of the total energy with respect to the prefect BCC structure as a function of the rhombohedral distortion magnitude $\delta$ at a pressure of 0 GPa for vanadium calculated with different xc functionals. Under this condition, the BCC phase of V should be stable against any distortions as observed in experiment. This physical requirement is qualitatively fulfilled by local and semi-local functionals. However, advanced hybrid functionals, including the widely used HSE06, PBE0, and B3LYP, all describe the instability of BCC vanadium wrongly, and predict a transition to $RH_1$ phase.

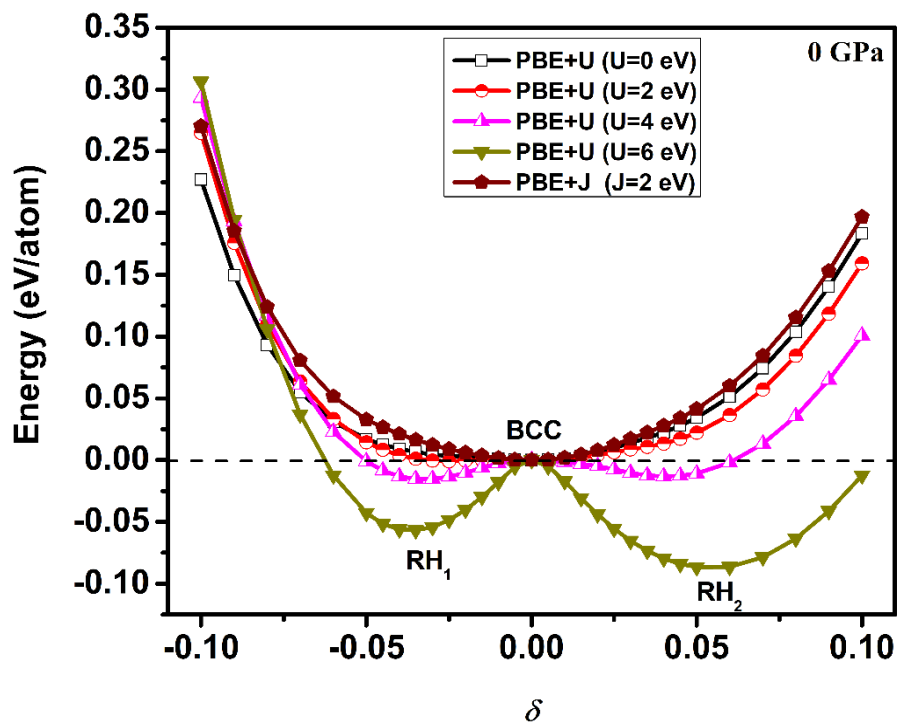

**Figure 3.** (Color online) Variation of the total energy with respect to the prefect BCC phase of vanadium at zero pressure as a function of the rhombohedral distortion magnitude $\delta$ calculated with different values of $U$ or $J$ parameter in the DFT+$U$ or DFT+$J$ method, respectively. The resemblance of the DFT+$U$ results with that of hybrid functionals suggests that its unphysical behavior (Fig. 2) is due to overestimation of the electron localization in the exchange part.

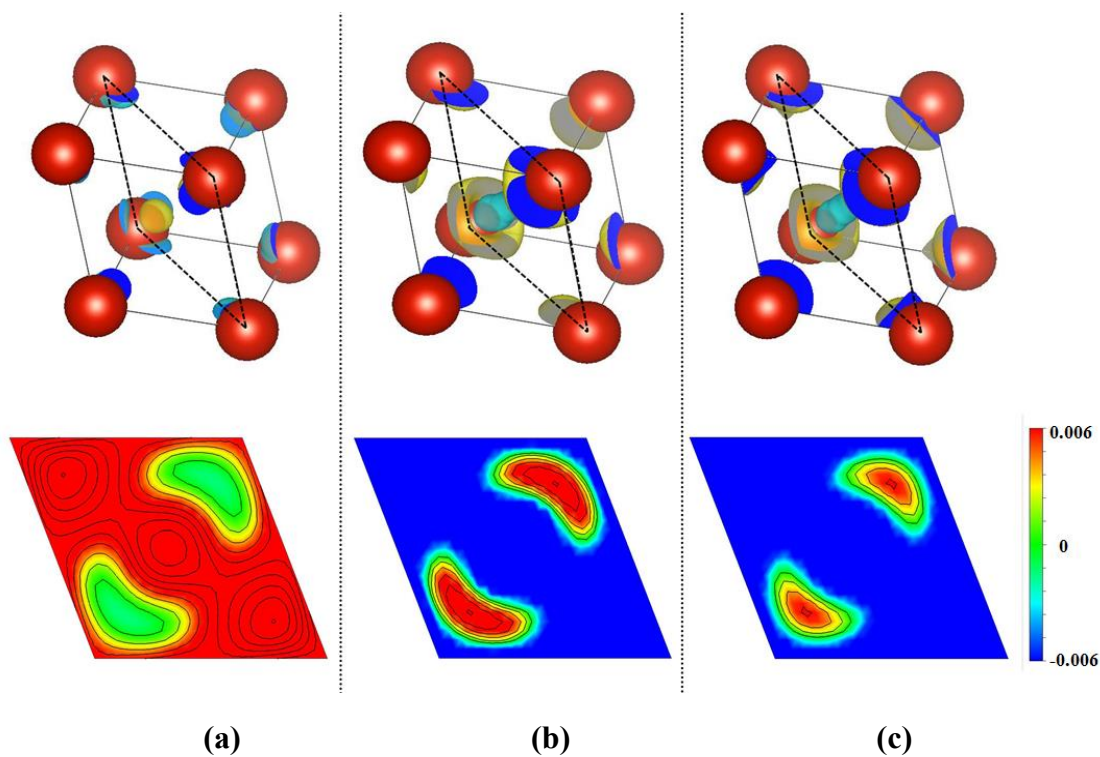

**Figure 4.** (Color online) The differential charge density (the color of yellow and blue indicate positive and negative value, respectively, with an isosurface level of 0.0202 e/Å$^3$) (top panel), and the differential electron localization function (ELF) in the given plane (indicated by the black dotted lines in the top panel) (bottom panel) with respect to that of PBE results, calculated with (a) PBE+$J$ with $J$=2 eV, (b) PBE+$U$ with $U$=4 eV, and (3) HSE06 method, respectively, for vanadium at zero pressure with a rhombohedral deformation of $\delta$=0.04. The similarity between DFT+$U$ and hybrid functional is evident. All have the same color map and contour values (with an interval of 0.002).

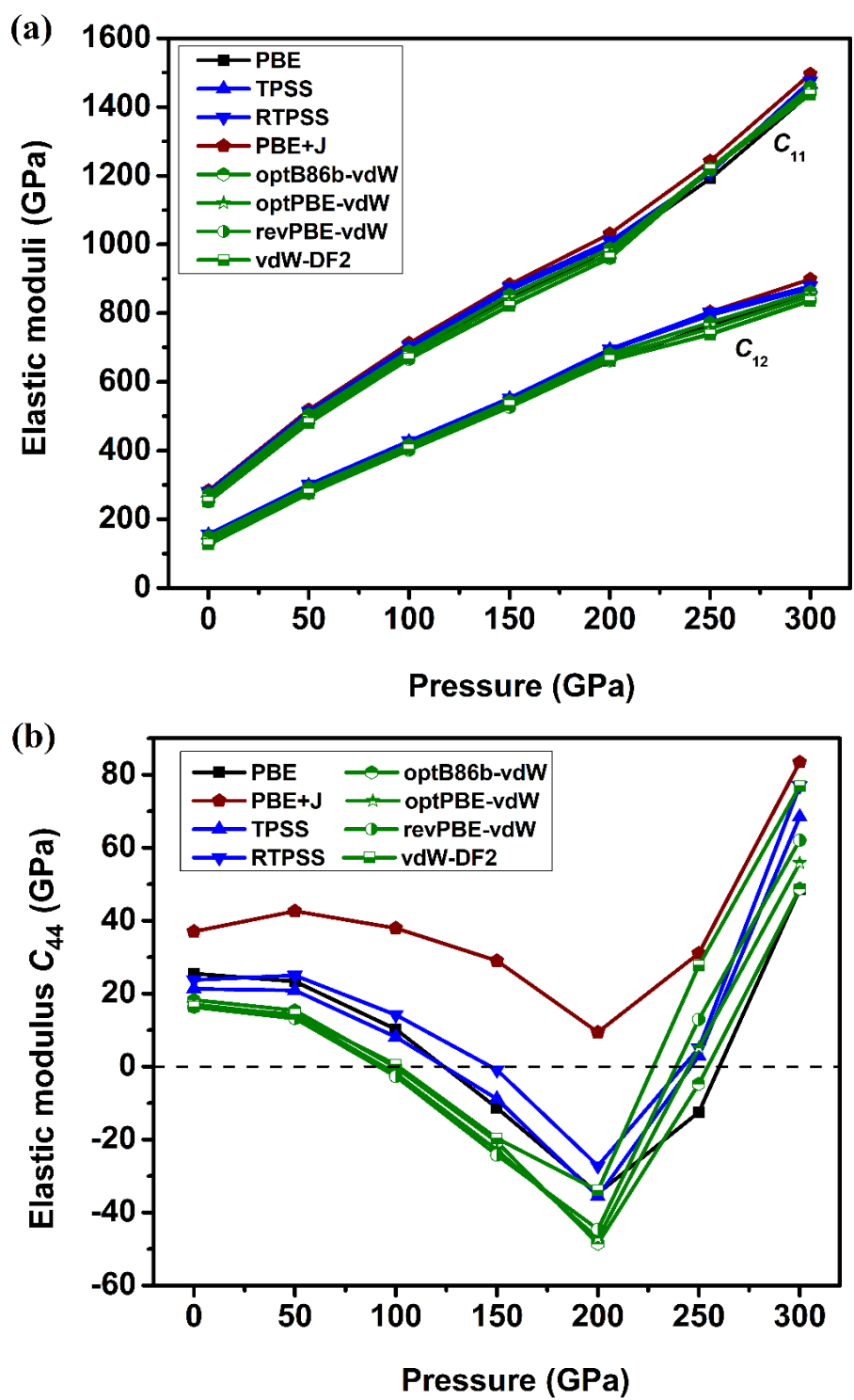

**Figure 5.** Calculated elastic moduli of BCC V as a function of pressure at zero Kelvin by using different exchange-correlation functionals and the PBE+$J$ method (with $J$=2 eV): (a) $C_{11}$ and $C_{12}$, and (b) $C_{44}$.

# Supplementary Material

# Orbital localization error of density functional theory in shear properties of vanadium and niobium

Yi X. Wang,[1,2] Hua Y. Geng,[1,*] Q. Wu,[1] and Xiang R. Chen[3]

[1]National Key Laboratory of Shock Wave and Detonation Physics, Institute of Fluid Physics, CAEP, Mianyang 621900, Sichuan, People's Republic of China

[2]College of Science, Xi'an University of Science and Technology, Xi'an 710054, People's Republic of China

[3]Institute of Atomic and Molecular Physics, College of Physical Science and Technology, Sichuan University, Chengdu 610064, China

## SI. Further details of computation

All simulations reported in the main text were computed using the Vienna *Ab initio* Simulation Package (VASP), which is based on density functional theory (DFT) and the projector augmented-wave (PAW) method. Since the local and semi-local functionals cannot quantitatively describes the shear properties of V and Nb, we also investigated other commonly used functionals such as hybrid functionals and DFT+$U$ formalism following the recipe of Liechtenstein *et al.*[72]

The following procedure was adopted to evaluate the EOS. First, for a given lattice constant $\varepsilon$, we calculate the total energy $E$ and the corresponding primitive cell volume $V$. Then the atomic structure in a supercell is fully optimized (to take the possible local distortion into account) to obtain the lowest energy $E_{min}$ for the given $V$. This procedure is repeated over a wide range of $\varepsilon$, the obtained $E_{min}$-$V$ data are then fitted to the Vinet equation of state[75]

$$\ln\left(\frac{Px^2}{3(1-x)}\right) = \ln B_0 + a(1-x), \tag{1}$$

in which

$$x = \left(\frac{V}{V_0}\right)^{1/3}, \quad a = \frac{3}{2}(B_0' - 1). \tag{2}$$

By taking the derivative of the E-V curve, we acquire the pressure-volume EOS.

To evaluate the elastic constants of the V and Nb, we use a conventional 2-atom cubic unit cell to calculate the total energy as a function of volume and the distortion along the shear strains. The shear moduli $C_{44}$ and $C' = (C_{11}-C_{12})/2$ are then obtained from the second derivatives of the total energy with respect to the deformation magnitude $\delta$, which is defined in the strain matrices as[42]

$$\varepsilon_{C_{44}} = \begin{pmatrix} 0 & \delta & 0 \\ \delta & 0 & 0 \\ 0 & 0 & \delta^2/(1-\delta^2) \end{pmatrix}, \tag{3}$$

$$\varepsilon_{C_{11}-C_{12}} = \begin{pmatrix} \delta & 0 & 0 \\ 0 & -\delta & 0 \\ 0 & 0 & \delta^2/(1-\delta^2) \end{pmatrix}. \tag{4}$$

The corresponding strained energies are expressed as

$$E(\delta) = E(0) + 2V C_{44} \delta^2 + O(\delta^3) \tag{5}$$

$$E(\delta) = E(0) + V(C_{11}-C_{12})\delta^2 + O(\delta^3) \tag{6}$$

The other two elastic constants $C_{11}$ and $C_{12}$ are then derived by using the calculated bulk modulus

$$B = -V\left(\frac{\partial P}{\partial V}\right)_T = \frac{1}{3}(C_{11} + 2C_{12}) \tag{7}$$

Besides the above-mentioned shear deformations, we also investigate the RH distortions, inspired by the BCC→RH structural transition observed in vanadium. The volume-conserved BCC→RH distortion matrix is defined as

$$T(\delta) = \begin{pmatrix} k & \delta & \delta \\ \delta & k & \delta \\ \delta & \delta & k \end{pmatrix}, \tag{8}$$

in which k is determined from the real positive solution of det(T) = 1, to ensure a volume-conserving transformation. The small displacement $\delta$ represents the amount of the rhombohedral deformation in the BCC crystalline system. The details of this distortion are referenced to Refs. 35 and 45.

**SII. Brief introduction to some typical functionals**

In order to assess the performance of density functional theory on transition metal V and Nb, different exchange-correlation functionals, such as GGA, hybrid functionals, PBE+$U$, meta-GGA, and van der Waals functionals, are carried out thoroughly in this work. These functionals differ in some fundamental issues: (i) GGA relies on an approximate treatment of the density gradient effect on the exchange-correlation energy; (ii) Hybrid functional includes a portion of non-local and fully orbital dependent exact exchange, plus the LDA or GGA correlation; (iii) PBE+$U$ contains the same PBE approximate exchange-correlation, but takes into account the orbital dependence (here applied to the $d$ states for V and Nb) of the Coulomb and exchange interactions which is absent in the PBE; (iv) Meta-GGA includes the second derivative of the electron density (the Laplacian), so it completes the third rung of "Jacob's ladder" of approximations, above the LDA and GGA rungs; (v) Van der Waals functionals are constructed by combining the nonlocal dispersion correlation energy with the exchange and local-correlation energies treated in usual LDA and GGA.

**(i) Hybrid functionals**

Hybrid functionals are characterized by mixing a portion of nonlocal Fock exchange with local or semi-local DFT exchange in a certain proportion. The construction of hybrid functional is motivated by the fact that the deficiencies of DFT and Hartree-Fock (HF) are in some sense complementary: band gaps predicted by DFT are too narrow (usually due to delocalization error), whereas the gaps calculated using HF are often too wide (usually due to localization error). Hence there is a hope that a mixed functional of them may not only predict more accurate gap, but also lead to more accurate total energy and geometry, because of the error cancellation. The mixing of HF and DFT exchange energies may also be justified to some extent by the adiabatic connection formula for the exchange energy.[76,77] The first and the most popular hybrid functional is the three-parameter B3LYP functional, which mixing 80% of LDA with 20% of HF exchange (adding a certain amount of Becke's correction, $\Delta E_x^{B88}$), and mixing in the correlation part 19% of the Vosko-Wilk-Nusair (VWN) functional[78] with 81% of Lee-Yang-Parr[79] correlation (note that these exchange and correlation functionals in turn contain some empirical parameters). The final functional is

$$E_{xc}^{B3LYP} = 0.8 E_x^{LDA} + 0.2 E_x^{HF} + 0.72 E_x^{B88} + 0.19 E_c^{VWN} + 0.81 E_c^{LYP} \qquad (9)$$

The mixing coefficients have been determined by fitting to a test set of molecules. The B3LYP functional is popular in molecular quantum chemistry. However, it has to be pointed out that the correlation part of the functional is incorrect in the limit of the homogeneous electron gas. It is found by Paier et al.[80] that the B3LYP hybrid functional successfully predicts a wide range of molecular properties. For periodic systems, however, the failure to attain the exact homogeneous electron gas limit as well as the semiempirical construction turns out to be a major drawback of this functional. This limitation, however, is irrelevant to its failure in V and Nb as reported in the main text.

An attempt to reduce the degree of empiricism has been made with the PBE0 and HSE06 functionals. The PBE0 functional mixes 1/4 of exact (HF) exchange with 3/4 of PBE exchange, and describes the correlation in the GGA (PBE),

$$E_{xc}^{PBE0} = 0.25 E_x^{HF} + 0.75 E_x^{PBE} + E_c^{PBE}. \qquad (10)$$

For molecular systems, the improvement achieved with the PBE0 functional is well documented, and it can be attributed to the fact that an admixture of a certain part of exact exchange reduces the self-interaction error in DFT.[81]

Under periodic boundary conditions, the calculation of the HF exchange energy is very expensive because of the slow decay of the exchange interaction with distance. To avoid this difficulty, Heyd et al.[65] proposed to separate the Coulomb kernel into a short- and long-range part,

$$\frac{1}{r} = S_\mu(r) + L_\mu(r) = \frac{erfc(\mu r)}{r} + erf(\mu r)r, \tag{11}$$

where $\mu$ is the range-separation parameter determining the distance (equal to $\sim 2/\mu$) beyond which the short-range interaction becomes negligible. In the HSE06 functional, the mixing of HF and DFT exchange is applied only to the short-range interaction, i.e.,

$$E_{xc}^{HSE} = 0.25 E_x^{sr}(\mu) + 0.75 E_x^{sr,PBE}(\mu) + E_x^{lr,PBE}(\mu) + E_c^{PBE}. \tag{12}$$

It has been established empirically that $\mu \sim 0.2$ to $0.3$ Å$^{-1}$ is a good universal choice for the range-separation parameter. The mixing ratio adopted in the PBE0 and HSE06 functional is not the result of a fitting. It has been argued[25] that this ratio is suggested by the result of a coupling constant integration between the HF and DFT limit.

In our simulations, we used a $14 \times 14 \times 14$ uniform mesh for the $k$-point sampling for the hybrid functionals HSE06, PBE0, and B3LYP. The internal energy differences with respect to the perfect BCC phase of vanadium at zero pressure as a function of the rhombohedral distortion $\delta$ calculated by using the HSE06 hybrid functional with different size of the $k$-point mesh are shown in the Fig. S5. It is evident that a $14 \times 14 \times 14$ uniform mesh is enough to give a well converged total energy.

### (ii) PBE+$U$ method

The DFT often fails to describe systems with localized and strongly correlated $d$ and $f$ electrons (usually with a narrow band). This manifests itself primarily in the form of unrealistic one-electron energies. In some cases, this can be remedied by introducing a strong intra-atomic interaction in a (screened) Hartree-Fock like manner, as an on-site replacement of the PBE. This approach is commonly known as the DFT+$U$ method.

The rotationally invariant DFT+$U$ introduced by Liechtenstein *et al*.[72] has a form of

$$E_{HF} = \frac{1}{2} \sum_{\{\gamma\}} (U_{\gamma_1\gamma_3\gamma_2\gamma_4} - U_{\gamma_1\gamma_3\gamma_4\gamma_2}) \hat{n}_{\gamma_1\gamma_2} \hat{n}_{\gamma_3\gamma_4}. \quad (13)$$

It is determined by the PAW on-site occupancies $\hat{n}$, and the (unscreened) on-site electron-electron interaction

$$U_{\gamma_1\gamma_3\gamma_2\gamma_4} = \left\langle m_1 m_3 \left| \frac{1}{\|r-r'\|} \right| m_2 m_4 \right\rangle \delta_{s_1 s_2} \delta_{s_3 s_4}, \quad (14)$$

in which $|m\rangle$ are the spherical harmonics. The unscreened e-e interaction $U_{\gamma_1\gamma_3\gamma_2\gamma_4}$ can be written in terms of Slater's integrals $F^0$, $F^2$, $F^4$, and $F^6$ ($f$-electrons). Using values for the Slater integrals calculated from atomic orbitals, however, would lead to a large overestimation of the true e-e interaction, since in solids the Coulomb interaction is screened (especially $F^0$). In practice, these integrals are therefore often treated as parameters, and adjusted to reach an agreement with experiment. The usually employed test set includes equilibrium volume, magnetic moment, band gap, and structural geometry. They are normally specified in terms of the effective on-site Coulomb- and exchange parameters $U$ and $J$.

The total energy functional in the DFT+$U$ formalism can be written as

$$E_{tot(n,\hat{n})} = E_{DFT}(n) + E_{HF}(\hat{n}) - E_{dc}(\hat{n}), \quad (15)$$

where the Hartree-Fock like interaction replaces the PBE on-site. One needs to subtract a double counting energy ($E_{dc}$) which supposedly equals the on-site PBE contribution to the total energy,

$$E_{dc}(\hat{n}) = \frac{U}{2} \hat{n}_{tot}(\hat{n}_{tot} - 1) - \frac{J}{2} \hat{n}_{tot}^{\sigma}(\hat{n}_{tot}^{\sigma} - 1). \quad (16)$$

In our case of V and Nb, we use PBE+$U$ technique to treat the (possible) on-site Coulomb interactions. We apply the method to the $d$ orbitals with the spin polarization being considered, to investigate the influence of $d$ electron localization on shear modulus and phase stability of BCC structure. We also employ the PBE+$J$ formalism (with $U$=0 eV) to enhance the on-site electron exchange, thus reduces the electron localization. This approach produces a result in good agreement with experimental data.

**(iii) Meta-GGA**

Based on the assumption of density functional, the energy functional can be expanded into a series of contributions that depend on density derivatives (similar to the Taylor expansion of a function), thus forming a Jacob's ladder[82] towards the exact functional. Taking higher rungs into account is a promising route to improve the exchange-correlation functionals. The first rung is the local density approximation (LDA), which depends only on local electron densities. If include not only densities but also their gradients, one gets the gradient approximation, such as the GGA,[83,84] which constitutes the second rung. Further addition of the second derivatives of the electron density (the Laplacian) leads to meta functionals, such as meta-GGA, which depend on the kinetic energy density and form the third rung. The most popular meta-GGA functionals include TPSS, RTPSS, M06L, and MBJ,[85-87] as well as others.

In 2011, Sun *et al.*[73] performed a self-consistent calculation within the framework of the PAW method to assess the performance of TPSS and RTPSS for the structural properties and thermochemical properties of solids and molecules. It is found that both TPSS and RTPSS yield accurate atomization energies for the molecules in the AE6 set, better than those of the standard PBE. For lattice constants and bulk moduli of 20 diverse solids, RTPSS performs much better than PBE. In addition, they also studied the magnetic properties of Fe, for which both TPSS and RTPSS predict the correct ferromagnetic ground-state solid phase, with an accurate magnetic moment.

Recently, Sun *et al.*[73] constructed SCAN meta-GGA that is fully constrained, obeying all 17 known exact constraints that a meta-GGA can. It is also exact or nearly exact for a set of "appropriate norms", including rare-gas atoms and nonbonded interactions. This strongly constrained and appropriately normed meta-GGA achieves remarkable accuracy for systems where the exact exchange-correlation hole is localized near its electron, and especially for lattice constants and weak interactions. Their further study[88] showed that SCAN is superior to the PBE for some standard molecular test sets and a small collection of solids. The mean absolute errors for SCAN are smaller than those for PBE by a factor of about 4 for the atomization energies of the 223 G3 molecules, a factor of 3 for the binding energies of the S22 set of weakly bound dimers of small molecules, and a factor of 4 for the LC20 set lattice constants for solids. SCAN is also more accurate, by about 30%, in predicting the BH76 energy barriers to chemical reactions. Studies also showed that the mean absolute errors of SCAN for the heats of formation of 94 binary solids are smaller than those of PBE by about 30%, or a factor of 3, for compounds with or without transition-metal elements, respectively. Moreover, SCAN has an unexpected and striking performance for diversely bonded systems, many of which were believed to be out of reach of semi-local functionals, and is comparable to or even more accurate than a computationally more expensive hybrid functionals.

In this work, we observed a gradual improvement of DFT in V and Nb from LDA, GGA to meta-GGA. Nonetheless, the correction is still short of capturing all localization and delocalization error in these metals. Especially, SCAN does not improve the result. In Nb, it is slightly worse than TPSS. This highlights the necessity to include higher order derivatives of density and move onto the next rung of the Jacob's ladder.

**(iv) Van der Waals functionals**

The van der Waals density functional (vdW-DF) was developed by Dion *et al.*[74] in 2004. In this approach, an approximation of interacting response functions in terms of electronic density was developed to evaluate the nonlocal correlation energy—the component of total energy that is responsible for long-range dispersion forces. Thus, similar to standard local or semilocal functionals, this nonlocal correlation energy can be expressed as a functional of electron density via a double integral where the electron density at two spatial points is correlated by a nonlocal kernel that depends on the density and density gradient at those two points. In fact, the vdW-DF by Dion *et al.* was constructed by combining the nonlocal correlation energy with the exchange and local-correlation energies treated in usual LDA and GGA. As a result, vdW-DF can be used just as other standard functionals in DFT electronic structure codes. Exchange functionals that follow this philosophy include optB86b-vdW, optPBE-vdW, revPBE-vdW, and vdW-DF2. Up to now, many studies have shown that the vdW-DFs can describe well for structural and energetic properties of various van der Waals complexes.[89-92] Qualitatively, van der Waals functionals enhance electron localization, and gives large binding energy and bigger band gap in insulators and semiconductors. In order to have a comprehensive comparison, we also employed these functionals to investigate their performance in V and Nb in this work. The results showed that the van der Waals interaction enhanced $d$ electron localization, which worsens the quality of vdW-DF results by comparison to GGA and meta-GGA. Its performance is more similar to hybrid functionals and DFT+$U$ method.

## SIII. Supplementary references

Phys Rev B **76**, 125112 (2007).

## SIV. Supporting table and figure

**Table SI.** Calculated shear modulus $C_{44}$ of V and Nb at the ambient pressure by using different van der Waals functionals and the meta-GGA+$J$ method ($J$=2 eV for V, and $J$=1 eV for Nb), together with the available experimental data.

| $C_{44}$ (GPa) | revPBE-vdW | optPBE-vdW | vdW-DF2 | TPSS+$J$ | RTPSS+$J$ | Expt. |
|---|---|---|---|---|---|---|
| V | 16.31 | 17.04 | 16.42 | 36.80 | 39.09 | 42.5-44[49-55] |
| Nb | 13.36 | 14.72 | 20.09 | 26.84 | 25.64 | 28.7[56] |

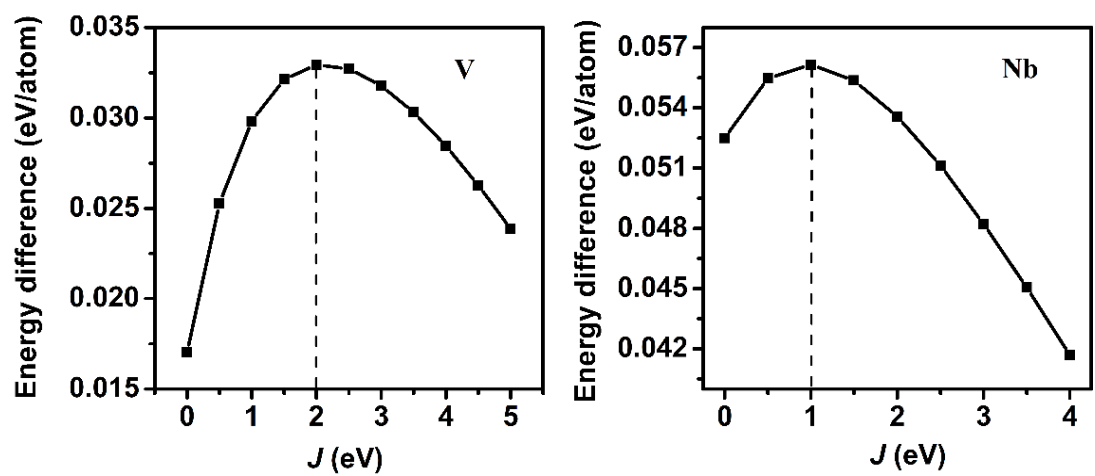

**Figure S1.** (Color online) Variation of the total energy difference at a rhombohedral deformation of $\delta=0.05$ with respect to the BCC phase as a function of the parameter $J$ acting on $d$ orbitals for V and Nb at zero pressure calculated using PBE+$J$ method. Note the maximum of these curves.

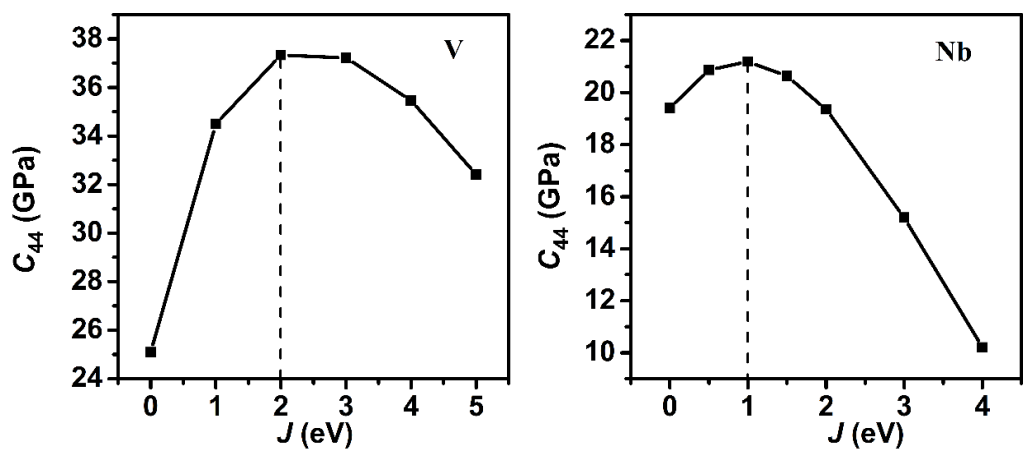

**Figure S2.** (Color online) Variation of shear modulus $C_{44}$ for vanadium and niobium at zero pressure calculated by PBE+$J$ method as a function of $J$ acting on the $d$ orbitals. Note the maximum of these curves.

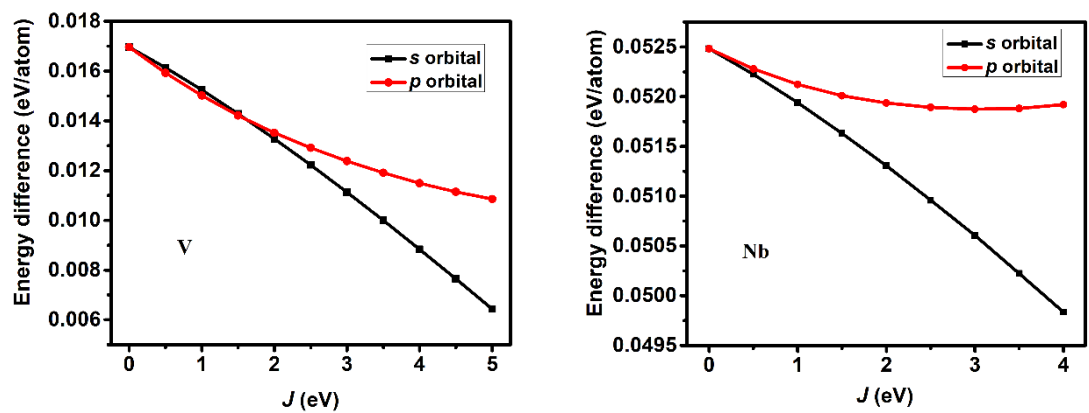

**Figure S3.** (Color online) Variation of the total energy difference at a rhombohedral deformation of $\delta=0.05$ with respect to the BCC phase for V and Nb at zero pressure calculated by PBE+$J$ method, with the parameter $J$ acting on $s$ and $p$ orbitals, respectively. The continuous decrease of the energy with respect to BCC reflects the delocalization error in $s$ and $p$ orbitals.

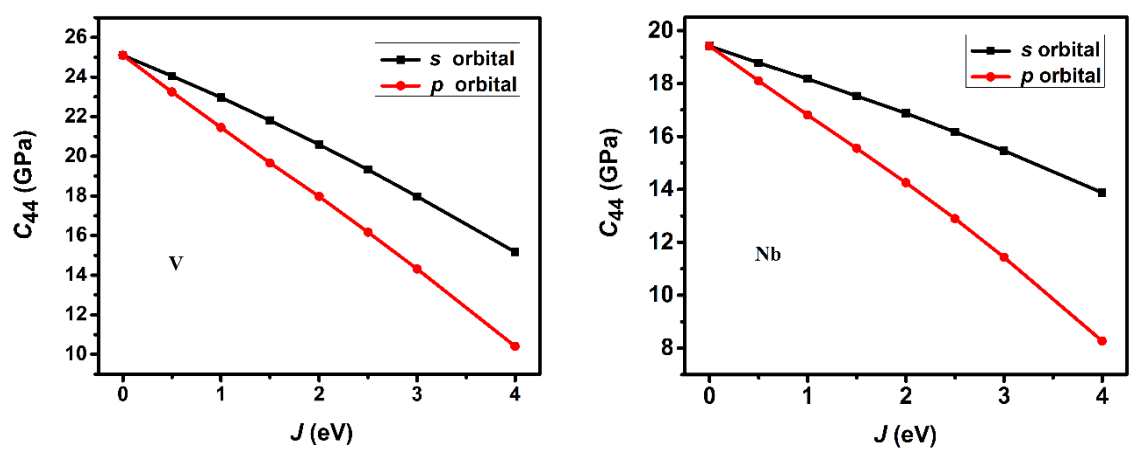

**Figure S4.** (Color online) Variation of shear modulus $C_{44}$ for vanadium and niobium at zero pressure calculated by PBE+$J$ method with $J$ acting on the $s$ and $p$ orbitals, respectively. The continuous decrease of $C_{44}$ reflects the impact of delocalization error from $s$ and $p$ orbitals.

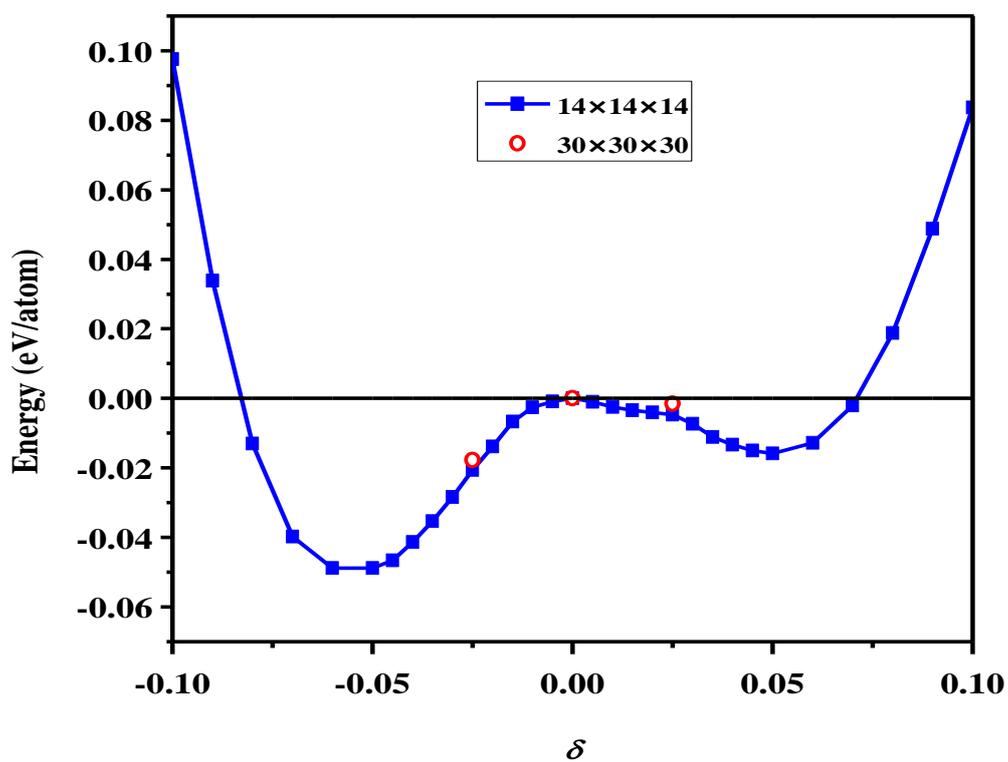

**Figure S5.** (Color online) The total energy differences with respect to the perfect BCC phase of vanadium at zero pressure as a function of the rhombohedral distortion $\delta$ calculated by using the HSE06 hybrid functional with different size of the *k*-points mesh. This result confirms the good convergency of the *k*-point size.

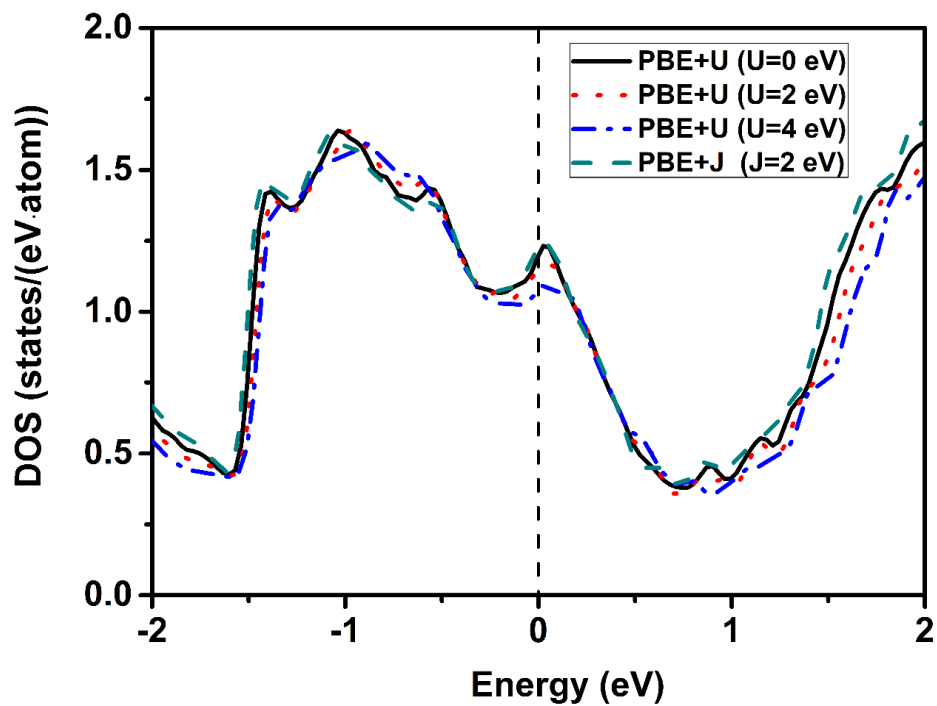

**Figure S6.** (Color online) Comparison of the electronic density of states (DOS) of vanadium at 0 GPa with a rhombohedral deformation of $\delta=0.04$ by using the PBE+$U$ and PBE+$J$ method.

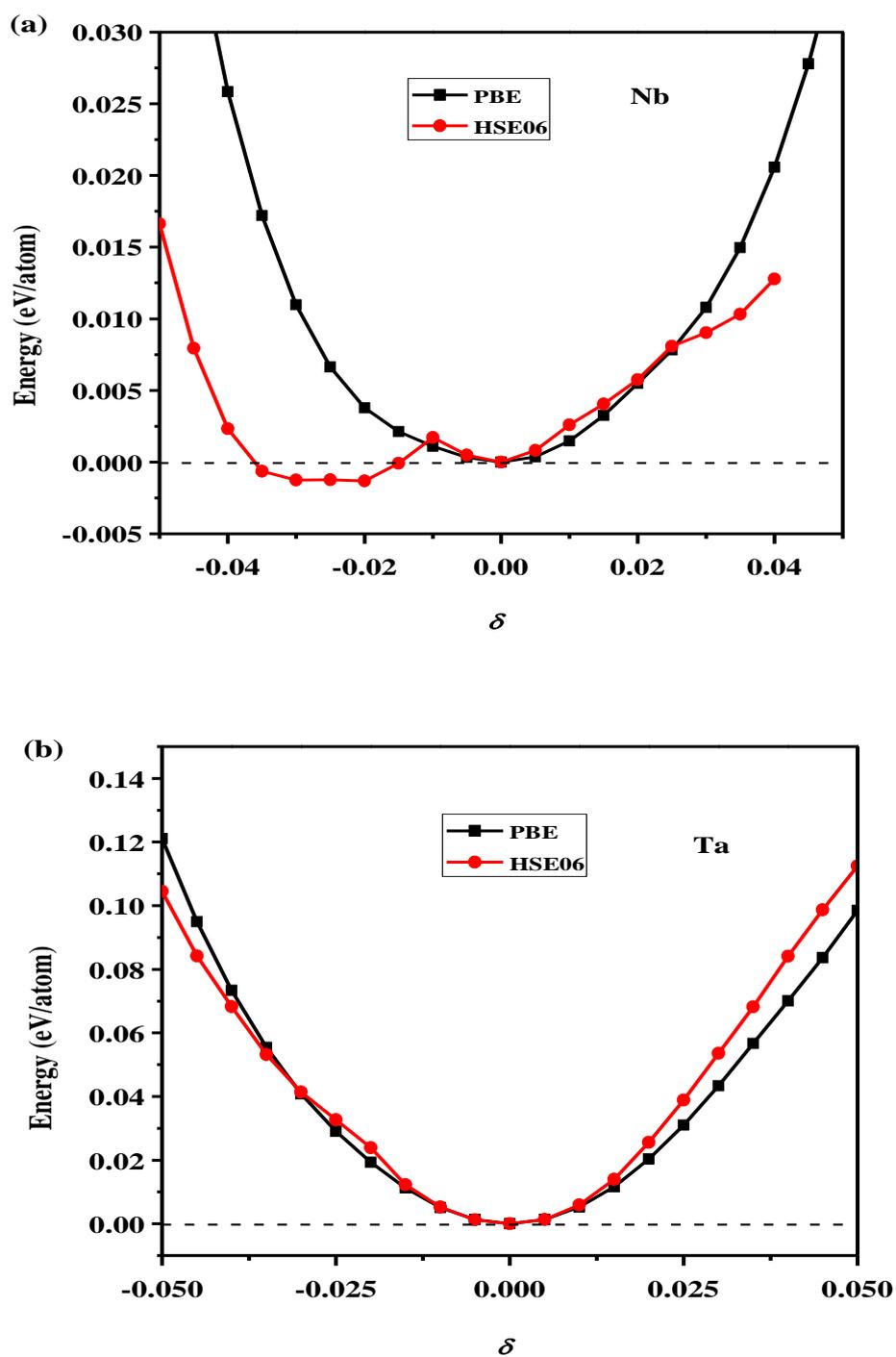

**Figure S7.** (Color online) The variation of total energy with respect to the prefect BCC phase as a function of the rhombohedral distortion $\delta$ at a pressure of 0 GPa for (a) niobium, and (b) tantalum, calculated with different exchange-correlation functionals.

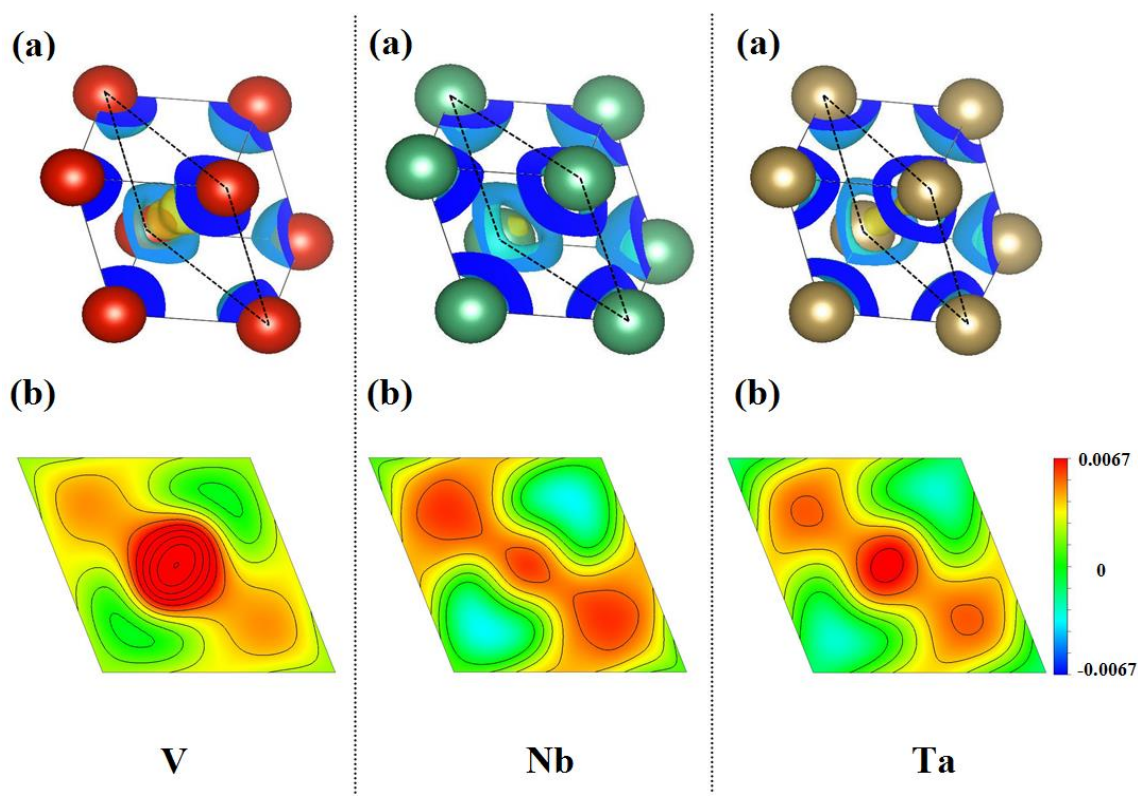

**Figure S8.** (Color online) The differential charge density isosurface (the color of yellow and blue indicate positive and negative value, respectively, with an isosurface level of 0.0067 e/Å$^3$) (top panel) and the corresponding contour map (bottom panel) on the given plane (indicated by the black dotted lines in the top panel) between PBE and PBE+$J$ (with $J$=2 eV) for V, Nb, and Ta with a rhombohedral deformation of $\delta$=0.04 at zero pressure, respectively. All have the same color map and contour values (the interval is 0.0013 e/Å$^3$).

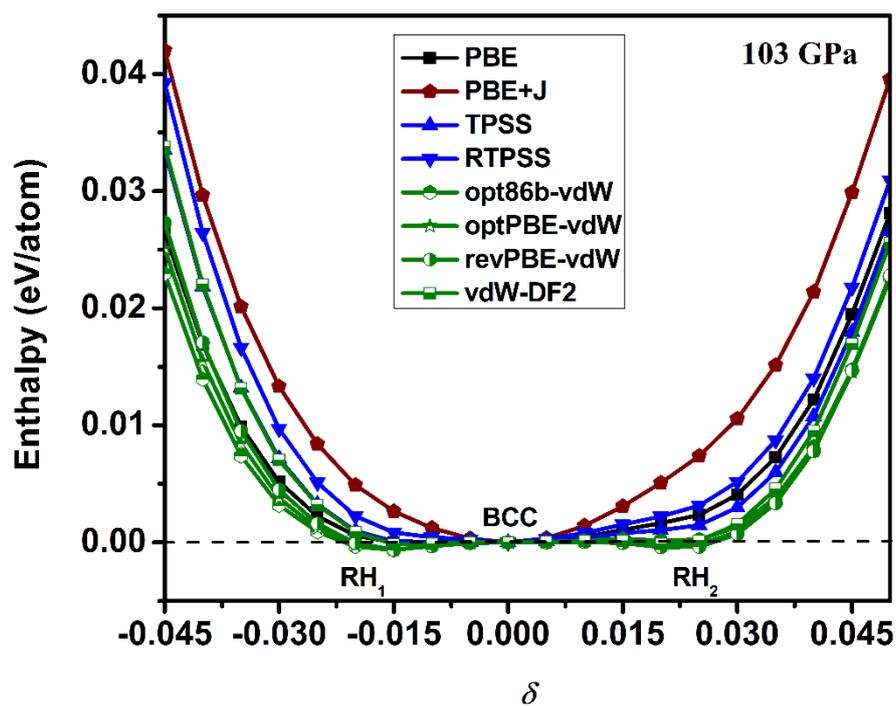

**Figure S9.** (Color online) Variation of the calculated enthalpy difference as a function of the rhombohedral deformation magnitude δ for vanadium at 103 GPa by using different exchange-correlation functionals and the PBE+*J* method. It is evident that correcting localization error enhances the stability of the BCC phase.